\documentclass[twocolumn,showpacs,preprintnumbers,amsmath,amssymb]{revtex4}
\usepackage{amsmath,amsfonts,diagbox,latexsym,amssymb,graphics,epsfig,subfigure,color,makeidx}
\usepackage{xcolor}
\usepackage[utf8]{inputenc}
\usepackage{graphicx,hyperref}
\usepackage{bm}
\usepackage{color}
\usepackage{multirow}

\begin{document}
	
	\title{Evolution of innermost stable circular orbit and light ring of a charged black hole induced by the scalarization}
	\author{Yu-Peng Zhang\footnote{zhangyupeng@lzu.edu.cn},
		Shao-Wen Wei\footnote{weishw@lzu.edu.cn},
		Yu-Xiao Liu\footnote{liuyx@lzu.edu.cn, corresponding author}
	}
	\affiliation{
	 Key Laboratory of Quantum Theory and Applications of MoE, Lanzhou Center for Theoretical Physics,\\
	 Key Laboratory of Theoretical Physics of Gansu Province, Gansu Provincial Research Center for Basic Disciplines of Quantum Physics, Lanzhou University, Lanzhou 730000, China\\
		Institute of Theoretical Physics \& Research Center of Gravitation, Lanzhou University, Lanzhou 730000, China
	}
	
	\begin{abstract}
		
In this paper, we investigate the dynamical evolution of the innermost stable circular orbit (ISCO) and light ring of a charged black hole during dynamical scalarization. This is achieved through nonlinear simulations within the framework of the Einstein-Maxwell-dilaton theory. Using the time-dependent metric derived from these simulations, we compute the radial effective potentials for timelike and null geodesics as functions of time. Our results demonstrate how the ISCO and light ring evolve as a hairless charged black hole transitions to a scalarized state. We find that dynamical scalarization induces an increase in the areal radii of both the ISCO and light ring. These findings provide new insights into black hole scalarization, particularly regarding the temporal evolution of the ISCO and light ring in a dynamically evolving spacetime.

	\end{abstract}
	\maketitle
	\section{Introduction}
	
Black holes are among the most intriguing compact objects predicted in the framework of general relativity. Observations of gravitational waves from binary black hole mergers \cite{Abbott2016a, LIGOScientific:2018jsj, LIGOScientific:2020kqk, LIGOScientific:2021psn} and direct imaging of black hole shadows \cite{eth2019, eth2022} have significantly advanced our ability to probe and characterize the dynamical properties of black holes in strong-field regimes. In Einstein gravity, it has been established that the final outcome of gravitational collapse for any form of matter is a Kerr-Newman black hole. This stationary, asymptotically flat solution is uniquely determined by three parameters: mass, charge, and angular momentum \cite{Chrusciel:2012jk}.

Chase initially demonstrated that in Einstein gravity, an asymptotically flat, static black hole cannot support minimally coupled massless scalar hair \cite{Chase:1970omy}. This result was extended by Hawking and Bekenstein, they proved that even minimally coupled, self-interacting static scalar fields still can not form scalar hair in the vicinity of stationary, asymptotically flat black holes \cite{Hawking:1972qk, Bekenstein:1971hc, Bekenstein:1972ky, Bekenstein:1995un}. These findings established a seemingly robust ``no-hair" theorem. However, this classical framework began to unravel with the exploration of novel field configurations. Specifically, the introduction of non-Abelian fields \cite{Volkov:1989fi, Bizon:1990sr, Greene:1992fw, Luckock:1986tr, Droz:1991cx}, non-minimally coupled fields \cite{Kanti:1995vq, Gregory:1992kr, Yamazaki:1989hy, Campbell:1991kz, Garfinkle:1990qj, Guo:2021zed, Yao:2021zid, Zhang:2021ybj, Zhang:2021nnn, Sotiriou:2013qea, Sotiriou:2014pfa, Benkel:2016kcq, Doneva:2017bvd, Silva:2017uqg, Andreou:2019ikc, Hod:2019pmb, Peng:2019snv, Ripley:2019aqj, Hod:2019vut, Dima:2020yac, Tang:2020sjs, Liu:2020yqa, Guo:2020sdu, Herdeiro:2020wei, East:2021bqk, Zhang:2022kbf, Wang:2020ohb}, and time-dependent oscillating complex scalar fields \cite{Herdeiro:2014goa, Sanchis-Gual:2015lje} in black hole spacetimes revealed scenarios where nontrivial ``hairy" solutions could persist, thereby challenging the original no-hair paradigm.

The linear analysis reveals that bosonic (scalar) fields exhibit instability, leading to exponential growth of the unstable modes over time \cite{Press:1972zz,Dolan:2007mj,Witek:2012tr,East:2017mrj,Bekenstein:1973ur, Hod:2012wmy,Sanchis-Gual:2015lje}. For instance, time-dependent scalar fields manifest superradiant instability in Kerr black hole spacetimes \cite{Press:1972zz, Dolan:2007mj, Witek:2012tr, East:2017mrj} or charged black hole backgrounds \cite{Bekenstein:1973ur, Hod:2012wmy, Sanchis-Gual:2015lje}. Another type of instability—tachyonic instability—can arise under non-minimal couplings, causing scalar fields to destabilize \cite{Silva:2017uqg, Doneva:2017bvd, Herdeiro:2018wub, Fernandes:2019rez, Dima:2020yac, Berti:2020kgk, Herdeiro:2020wei}. When accounting for scalar field back-reaction and energy conservation, black holes undergoing superradiant or tachyonic instabilities evolve into two distinct end states: synchronized configurations (for superradiant instabilities), and
scalarized hairy black holes (for tachyonic instabilities) \cite{Herdeiro:2014goa, Sanchis-Gual:2015lje, East:2017ovw, Silva:2017uqg, Doneva:2017bvd, Herdeiro:2018wub, Fernandes:2019rez, Dima:2020yac, Berti:2020kgk, Herdeiro:2020wei}. This dynamic evolution presents well-defined dynamical spacetimes with distinct initial and final states \cite{Sanchis-Gual:2015lje, Zhang:2021ybj, Zhang:2021nnn, lzqzyb:2023, Benkel:2016kcq, Zhang:2022kbf, Julie:2020vov, Witek:2020uzz, East:2020hgw, Doneva:2021tvn, East:2021bqk,Doneva:2020nbb, Doneva:2021dcc, Doneva:2021dqn, Witek:2018dmd, Benkel:2016rlz}. The transformation induces geometric modifications to the background spacetime, thereby influencing the innermost stable circular orbit (ISCO) and light ring structures.

The innermost stable circular orbit (ISCO) defines the inner boundary of an accretion disk surrounding a black hole \cite{Abramowicz:2011xu} and marks the threshold between stable and unstable circular orbits \cite{Cardoso:2008bp}. Equally pivotal is the light ring (also termed the photon sphere), which plays an essential role in imaging black hole shadows \cite{Luminet:1979nyg, Falcke:1999pj}. This light ring corresponds to bound null geodesics (both stable and unstable) and contributes to the black hole's shadow by delineating a region of enhanced brightness, arising from light rays that orbit the black hole multiple times before escaping to infinity \cite{Gralla:2019xty}.

Investigating time-dependent spacetime geometries can reveal how the ISCO and light ring evolve dynamically, offering insights into accretion disk dynamics and orbital stability in such systems \cite{Mishra:2019trb, Song:2021ziq}. To address this, we focus on a charged black hole undergoing dynamical scalarization, a process that generates time-evolving spacetime geometries. Using our spherical numerical relativity code \cite{Zhang:2023qag}, we analyze the temporal evolution of the ISCO and light ring in these dynamical spacetimes. Specifically, we construct a time-dependent effective potential following the methodology outlined in Ref. \cite{Cunha:2022gde} to track the evolution of both the ISCO and light ring. This approach enables precise characterization of their geometric properties as the spacetime evolves.

This paper is structured as follows. In Section~\ref{sec:fundamentals}, we introduce the Einstein-Maxwell-dilaton gravity framework and derive the coupled equations of motion governing gravitational and matter fields. Numerical results and their analysis are presented in Section~\ref{sec:results}. Finally, we give a brief conclusion and outlook in Sec.~\ref{Conclusion}.
	
\section{Model of dynamical spacetime} \label{sec:fundamentals}

To study the dynamical scalarization of a charged black hole in dynamical spacetime, we first introduce the theoretical framework described by the following action \cite{Garfinkle:1990qj,Zhang:2021ybj}:
\begin{equation}
S = \int d^4x \sqrt{|g|} \left( \frac{R - F(\phi)\mathcal{I}}{16\pi} - \frac{1}{2} \partial_\mu\phi \partial^\mu\phi - V(\phi) \right),
\label{action}
\end{equation}
where the coupling function is chosen as
\begin{equation}
F(\phi) = e^{\eta \phi},
\label{fofphi}
\end{equation}
with $\eta$ denoting the scalar-electromagnetic coupling parameter. The electromagnetic invariant $\mathcal{I}$ is constructed via the Faraday tensor $F_{\alpha\beta} = \nabla_\alpha A_\beta - \nabla_\beta A_\alpha$ as $\mathcal{I} = F_{\alpha\beta}F^{\alpha\beta}$. It is important to note that the choice of $F(\phi)$ critically determines the existence of scalarized/un-scalarized black holes: specifically, the condition $\frac{dF(\phi)}{d\phi} = 0$ permits static un-scalarized solutions \cite{Herdeiro:2018wub,Fernandes:2019rez}. Since our exponential coupling in Eq.~\eqref{fofphi} satisfies $\frac{dF(\phi)}{d\phi} \neq 0$, no static un-scalarized black holes exist in this model; however, the un-scalarized (hairless) Reissner-Nordström-like black hole still serves as a valid initial configuration for studying dynamical scalarization processes \cite{Sanchis-Gual:2015lje}. Throughout this paper, we adopt geometrized units with $G = c = \hbar = 1$ for simplicity.

Varying the action \eqref{action} with respect to the metric $g_{\mu\nu}$, electromagnetic potential $A_\mu$, and scalar field $\phi$ yields the coupled equations of motion:
\begin{eqnarray}
R_{\mu\nu} - \frac{1}{2}g_{\mu\nu}R &=& 8\pi \left(T_{\mu\nu}^{(\phi)} + T_{\mu\nu}^{(\text{em})}\right), \label{einsteineq}\\
\nabla_\mu\left(e^{\eta \phi}F^{\mu\nu}\right) &=& 0, \label{maxwelleq}\\
\nabla_\mu\nabla^\mu\phi &=& \frac{\eta}{16\pi}\mathcal{I}e^{\eta \phi}, \label{keleineq}
\end{eqnarray}
where the energy-momentum tensors $T_{\mu\nu}^{(\phi)}$ (scalar) and $T_{\mu\nu}^{(\text{em})}$ (electromagnetic) are explicitly defined as:
\begin{eqnarray}
	T^{(\phi)}_{\mu\nu} &=& \partial_\mu\phi\partial_\nu\phi - g_{\mu\nu}\left(\frac{1}{2}\partial^\alpha\phi\partial_\alpha\phi\right),\\
	T_{\mu\nu}^{(\text{em})} &=& \frac{1}{4\pi}e^{\eta \phi}\left(F_{\mu\alpha}F_\nu^{~\alpha}-\frac{1}{4}g_{\mu\nu}F^{\alpha\beta}F_{\alpha\beta}\right).
\end{eqnarray}
Notably, in the limit $\eta=0$, this system reduces to the standard Einstein-Maxwell theory minimally coupled to a scalar field, demonstrating the consistency of our framework with established physics.

The nonlinear dynamics of the system governed by Eqs. \eqref{einsteineq}, \eqref{maxwelleq}, and \eqref{keleineq} were simulated using our specialized spherical-coordinate numerical relativity code \cite{Zhang:2023qag,Zhang:2024wci}, which implements the Baumgarte–Shapiro–Shibata–Nakamura (BSSN) formalism~\cite{Montero:2012yr}. This code exhibits third-order convergence, with its algorithmic framework and validation extensively documented in Ref. \cite{Zhang:2023qag}.

The influence of scalar field self-interactions on dynamical scalarization phenomena was previously examined in Ref. \cite{Zhang:2024wci}. In this study, we initiate our analysis by focusing on the dynamical scalarization of a spherically symmetric charged black hole, deliberately neglecting scalar field self-interactions to simplify the initial computational framework. The spacetime geometry is described by the metric:
	\begin{eqnarray}
	ds^2&=&(-\alpha^2 + \beta^r \beta_r) dt^2 + 2\beta_r dt dr \nonumber\\
	&&+e^{4\chi}\left(a\,dr^2+b\,r^2 \,d\Omega^2 \right)
	\label{metric}
	\end{eqnarray}
where $d\Omega^2 = d\theta^2 + \sin^2\theta d\varphi^2$ denotes the angular component. Here, $\alpha$ (lapse function), $\beta^i = (\beta^r, 0, 0)$ (shift vector), $\chi$ (conformal factor), and $a$, $b$ (metric functions) are all functions of the radial coordinate $r$ and time $t$. The spatial 3-metric $\gamma_{ij}$ on the spacelike hypersurface $\Sigma_t$ is derived as:
\begin{equation}
\gamma_{ij} = e^{4\chi} , \mathrm{diag}\left(a, , b r^2, , b r^2 \sin^2\theta\right).
\end{equation}
For brevity, the evolution equations governing the metric functions, scalar, and electromagnetic fields are not explicitly presented here; their derivations are systematically outlined in Ref. \cite{Montero:2012yr}.

The scalar field is initially absent, and the spacetime is fully characterized by a charged black hole with mass $M_0$ and charge $Q$ at the initial time \cite{Sanchis-Gual:2015lje}. In this work, we investigate the evolution of the ISCO and the light ring driven by the scalarization process. To simplify our analysis, we consider a specific scenario involving a black hole with $M_0 = 1$ and $Q = 0.9$ at the initial time. Here, the conformal electric potential $\varphi$ and conformal factor $e^{\chi}$ are defined as \cite{Alcubierre:2009ij}:
	\begin{eqnarray}
	\varphi&=&\frac{Q}{r},\\
	e^{2\chi}&=& \left(1+\frac{M_0}{2r}\right)^2-\frac{Q^2}{4^2}.
	\end{eqnarray}
The corresponding physical electric field strength is
\begin{equation}
E^r = \frac{Q}{r^2 e^{6\chi}}. \label{eq:E_field}
\end{equation}
Under this configuration, the metric functions $a$ and $b$ are initialized as $a = b = 1$ at the initial time.

Next, we derive the differential equations governing the dynamical evolution of the ISCO and the light ring in the dynamical spacetime described by the metric \eqref{metric}. For an equilibrium spacetime with finite Arnowitt-Deser-Misner mass $M$ that is either spherically symmetric or axially symmetric and asymptotically flat, there exist two conserved quantities associated with the Killing vectors $(\partial_t)^\mu$ (energy $E$) and $(\partial_\varphi)^\mu$ (angular momentum $L_z$). By combining these conserved quantities with the geodesic equations of motion, one can construct a radial effective potential $V(r)$ to formulate the radial geodesic equation for timelike (ISCO) or null (light ring) trajectories. The ISCO and light ring correspond to the extrema of $V(r)$, which can be determined by solving the appropriate conditions on its derivatives.

However, the spacetime described by the metric \eqref{metric} is time-dependent and lacks a timelike Killing vector $(\partial_t)^\mu$. To address this issue, an adiabatic effective potential was proposed \cite{Cunha:2022gde}. It is critical to note that the background spacetime evolves rapidly, rendering the adiabatic approximation invalid. Despite this, we can still adopt a similar approach to define a time-evolving effective potential. To derive the time-dependent effective potential, we introduce the quasi-conserved energy $E$ and orbital angular momentum $J$ as follows:
\begin{eqnarray}
-E&=&g_{tt}u^t+g_{rt}u^r,\label{consE}\\
J&=&g_{\varphi\varphi}u^\varphi.\label{consJ}
\end{eqnarray}
Here, $u^r$ and $u^\varphi$ denote the radial and angular components of the four-velocity.

Because we are focusing on the ISCO and the light ring, we restrict our analysis to equatorial orbits with $u^\theta = 0$. For a test particle moving along a geodesic orbit, its four-velocity satisfies the normalization condition:
\begin{equation}
u^\mu u_\mu = c, \label{umu2}
\end{equation}
where $c=0$ corresponds to null geodesics (light rays) and $c=-1$ to timelike geodesics (massive particles). For timelike geodesics of massive particles, we set the mass to $m=1$. By combining Eqs. \eqref{consE}, \eqref{consJ}, and \eqref{umu2}, the radial component of the four-velocity is given by
\begin{equation}
\left(u^r\right)^2 = \frac{E^2 g_{\varphi\varphi} + J^2 g_{tt} - c, g_{tt}g_{\varphi\varphi}}{g_{\varphi\varphi}\left(g_{rt}^2-g_{rr}g_{tt}\right),} \label{effecV1}
\end{equation}
where the metric components are defined as:
\begin{eqnarray}
g_{tt} &=& -\alpha^2 + \beta^r\beta_r, \\
g_{rt} &=& \beta_r, \\
g_{\varphi\varphi} &=& e^{4\chi} b r^2 \sin^2(\theta).
\end{eqnarray}

We can recast Eq. \eqref{effecV1} into the form of an effective potential:
\begin{equation}
\left(u^r\right)^2=\frac{1}{g_{rt}^2-g_{rr}g_{tt}}\left( E^2 -V_{\text{eff}}\right),
\label{effecV2}
\end{equation}
where the time-dependent effective potential $V_{\text{eff}}$ is defined as:
\begin{equation}
V_{\text{eff}}(t,r,J)=\frac{c\,g_{\varphi\varphi}g_{tt}-J^2\,g_{tt}}{g_{\varphi\varphi}}.
\label{effpotential}
\end{equation}
For timelike geodesics ($c=-1$), the effective potential becomes:
\begin{equation}
V_{\text{eff}}(t,r,J)=\frac{-\,g_{\varphi\varphi}g_{tt}-J^2\,g_{tt}}{g_{\varphi\varphi}}.
\label{timelikeeffpotential}
\end{equation}
For null geodesics ($c=0$), the effective potential simplifies to:
\begin{equation}
V_{\text{eff}}(t,r,J)=\frac{-J^2\,g_{tt}}{g_{\varphi\varphi}}.
\label{nulleffpotential}
\end{equation}
Finally, we introduce the impact parameter $L=J/E$ to characterize the orbital motion of light, defined as the ratio of the orbital angular momentum $J$ to the energy $E$.

Using the derived effective potentials \eqref{timelikeeffpotential} and \eqref{nulleffpotential}, we can determine the ISCO for timelike geodesics and the light ring for null geodesics. For a black hole's ISCO, the following three conditions must be satisfied:

(a) The radial velocity of the geodesic vanishes:
\begin{equation}
u^r=0\label{condition1}.
\end{equation}

(b) The effective potential has an extremum (first derivative vanishes) and the acceleration of the radial motion is zero:
\begin{equation}
\frac{d V_{\text{eff}}}{d r}=0\label{condition2}.
\end{equation}

(c) The effective potential has a degenerate extremum (second derivative also vanishes), meaning the maximum and minimum of the potential coincide:
\begin{equation}
\frac{d^2V_{\text{eff}}}{dr^2}=0\label{condition3}.
\end{equation}

For the light ring, only the first two conditions (a) and (b) are required.

\section{Results} \label{sec:results}

Having established the effective potentials for timelike \eqref{timelikeeffpotential} and null \eqref{nulleffpotential} geodesics, and derived the conditions for determining the ISCO and light ring, we now proceed to investigate their dynamical evolution. Specifically, we perform numerical simulations of dynamical scalarization in charged black holes to study how the ISCO and light ring evolve in time-dependent spacetime geometries governed by our theoretical framework \eqref{action}. This model reveals that the properties of dynamical scalarization in charged black holes are determined by two key parameters: the black hole charge $Q$ and the coupling constant $\eta$. It has been shown that charged black holes with distinct values of $\eta$ relax into distinct scalarized configurations. Notably, the coupling parameter $\eta$ carries dimensions inverse to those of the scalar field $\phi$, such that the product $\phi\eta$ remains dimensionless.

In our previous work \cite{Zhang:2024wci}, we explored the dynamical scalarization of charged black holes under an axionic potential for the scalar field. For simplicity, we neglect the scalar field's self-interaction term and fixed the black hole charge at $Q = 0.9$. We then investigate the influence of the parameter $\eta$ on the dynamical evolution of the ISCO and light ring by considering the following cases with
\begin{equation}
\eta \phi_0 = (0, 5, 10, 15, 20),
\end{equation}
where $\eta \phi_0$ is dimensionless and we set $\phi_0 = 1$.

For a scalarized black hole described by the action \eqref{action}, the properties of such a black hole are governed by three key parameters: the electric charge $Q$, the coupling constant $\eta$, and the scalar field value at the apparent horizon $\phi_{\text{ah}}$. It is crucial to emphasize that numerical simulations of dynamical scalarization in charged black hole systems inherently require a coordinate transformation. The initial spacetime configuration is prescribed in isotropic coordinates with a vanishing shift vector. As the spacetime evolves dynamically, the shift vector evolves to counteract the spatial stretching of the spacelike hypersurface, gradually reaching a nearly stationary state at late times \cite{Alcubierre:2011pkc}.

We investigate the temporal evolution of the shift vector and metric components by numerically evolving a charged black hole spacetime while neglecting scalarization effects. Our analysis confirms that after an elapsed time of $t \approx 80 M_0$, the system transitions from its initial quasi-isotropic coordinate configuration into a new steady-state regime characterized by a non-vanishing shift vector. Notably, despite the unaltered physical properties of the black hole (i.e., mass, charge, and horizon scalar field value), the coordinate-dependent transformations in the metric structure and coordinate system induce modifications to the morphology of the effective potential.

To isolate the influence of black hole scalarization on the spacetime geometry, we suppress scalar field effects in all simulations during the initial time interval $t \leq 80 M_0$. This methodology prevents coordinate system artifacts arising from the shift vector's transition from zero to non-vanishing values in the quasi-isotropic coordinates initialization. Consequently, scalarization dynamics are only permitted to emerge once the simulation time surpasses $t = 80 M_0$.

\begin{figure*}[htbp]
		\includegraphics[width=\linewidth]{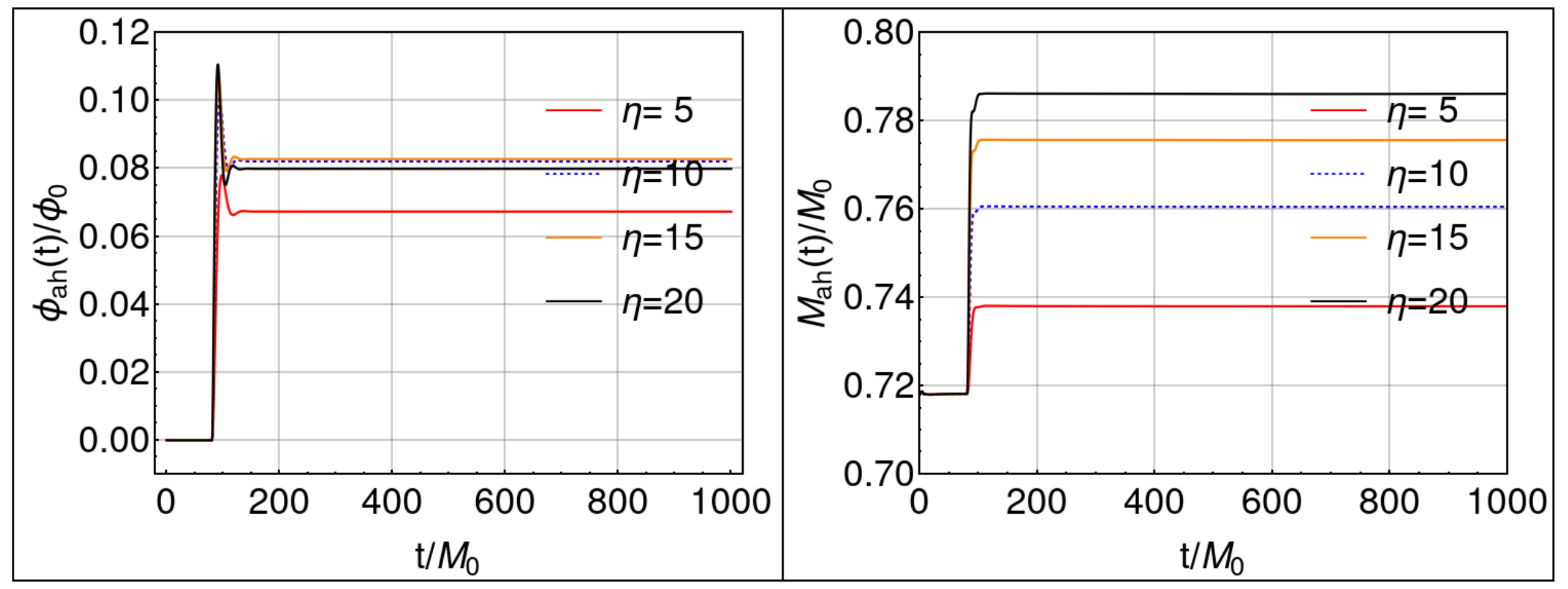}
	\caption{Time evolution of the scalar field on the apparent horizon and the black hole's irreducible mass $M_{\text{ah}}$ for varying coupling parameters $\eta$.}
	\label{p_m_phi_ah}
    \end{figure*}

\begin{figure*}[htbp]
		\includegraphics[width=\linewidth]{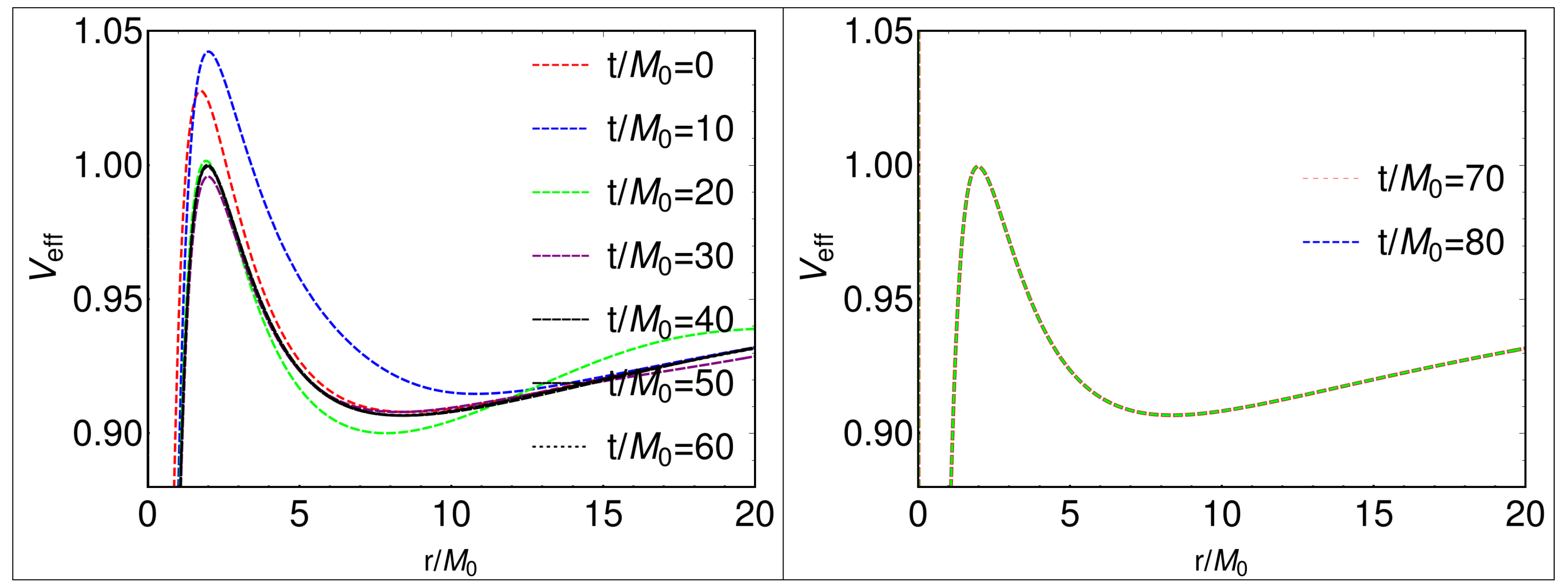}
	\caption{Time evolution of the effective potential through coordinate transformations for timelike geodesics with $\eta=5$ and $J=3.5$.}
	\label{time_veff_timelike}
    \end{figure*}

\begin{figure*}[htbp]
		\includegraphics[width=\linewidth]{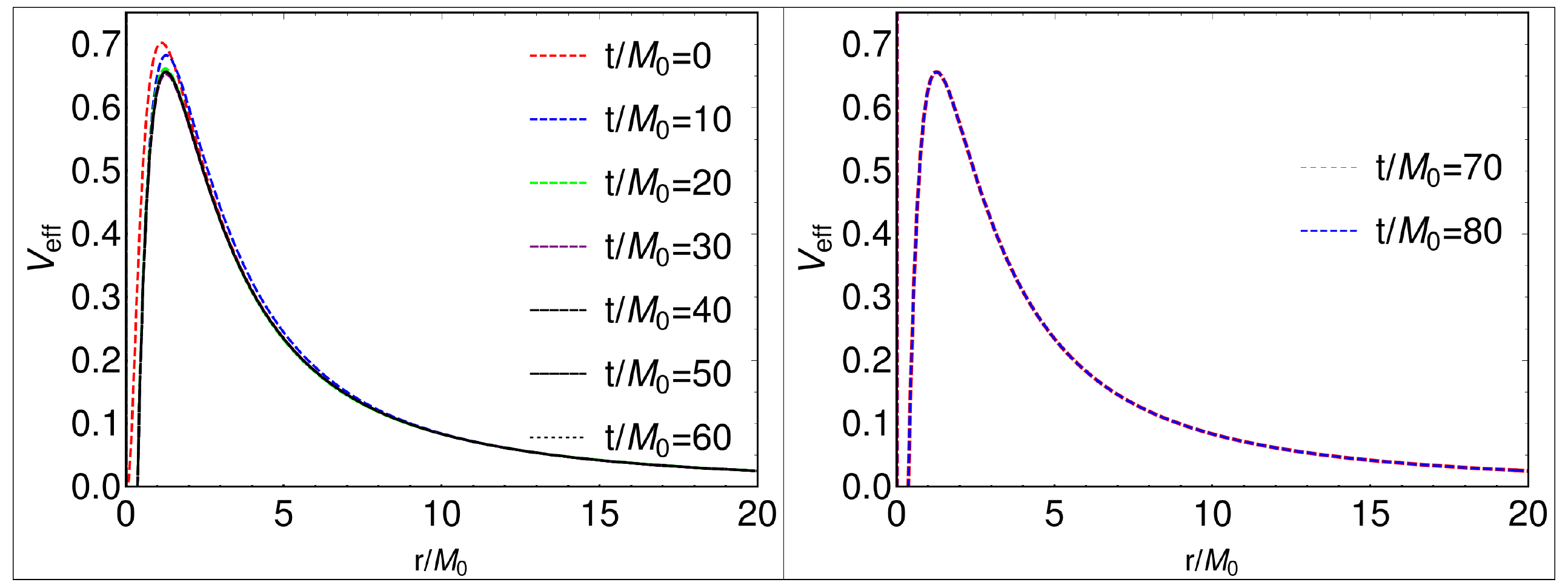}
	\caption{Time evolution of the effective potential through coordinate transformations for the null geodesics with $\eta=5$ and $L=3.5$.}
	\label{time_veff_null}
    \end{figure*}

The results of the dynamical scalarization process are presented in Fig. \ref{p_m_phi_ah}. As demonstrated in the figure, while the irreducible mass and horizon scalar field magnitude remain invariant, coordinate system evolution induces measurable alterations to the effective potentials defined in Eqs. \eqref{nulleffpotential} and \eqref{timelikeeffpotential}. Figures \ref{time_veff_timelike} and \ref{time_veff_null} illustrate the temporal evolution of these effective potentials during the coordinate transformation phase prior to $t = 80 M_0$. Notably, the effective potential configurations converge toward stable asymptotic states as the system approaches its quasi-stationary regime.

Once the spacetime evolves into a quasi-stationary regime, we introduce non-vanishing values for the coupling parameter $\eta$, and the charged black hole begins to transform into the scalarizing state. We have shown how scalarization induces changes in the irreducible mass and scalar field value at the apparent horizon in Fig. \ref{p_m_phi_ah}. Consequently, the background spacetime evolves over time as scalarization of the black hole progresses. We demonstrate that a charged black hole will reach different final states depending on the coupling parameter $\eta$. These differences imply that the ISCO and the light ring will evolve in distinct ways. For different coupling parameters with $\eta = (5, 10, 15, 20)$, the effective potential displays qualitatively similar evolutionary behavior during temporal evolution, but exhibits distinct quantitative variations in specific aspects.

\begin{figure*}[htbp]
		\includegraphics[width=\linewidth]{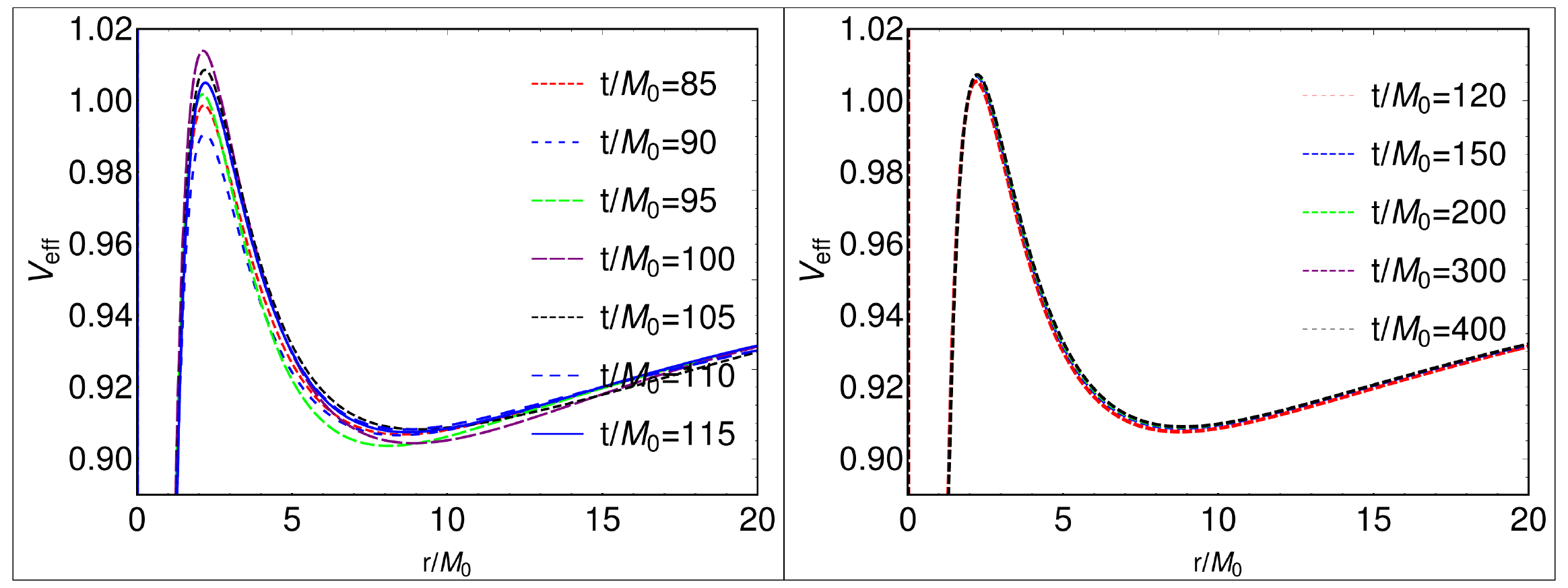}
	\caption{Time evolution of the effective potential for the timelike geodesics with $\eta=5$ and $J=3.5$.}
	\label{time_veff_timelike_evo}
\end{figure*}

\begin{figure*}[htbp]
		\includegraphics[width=\linewidth]{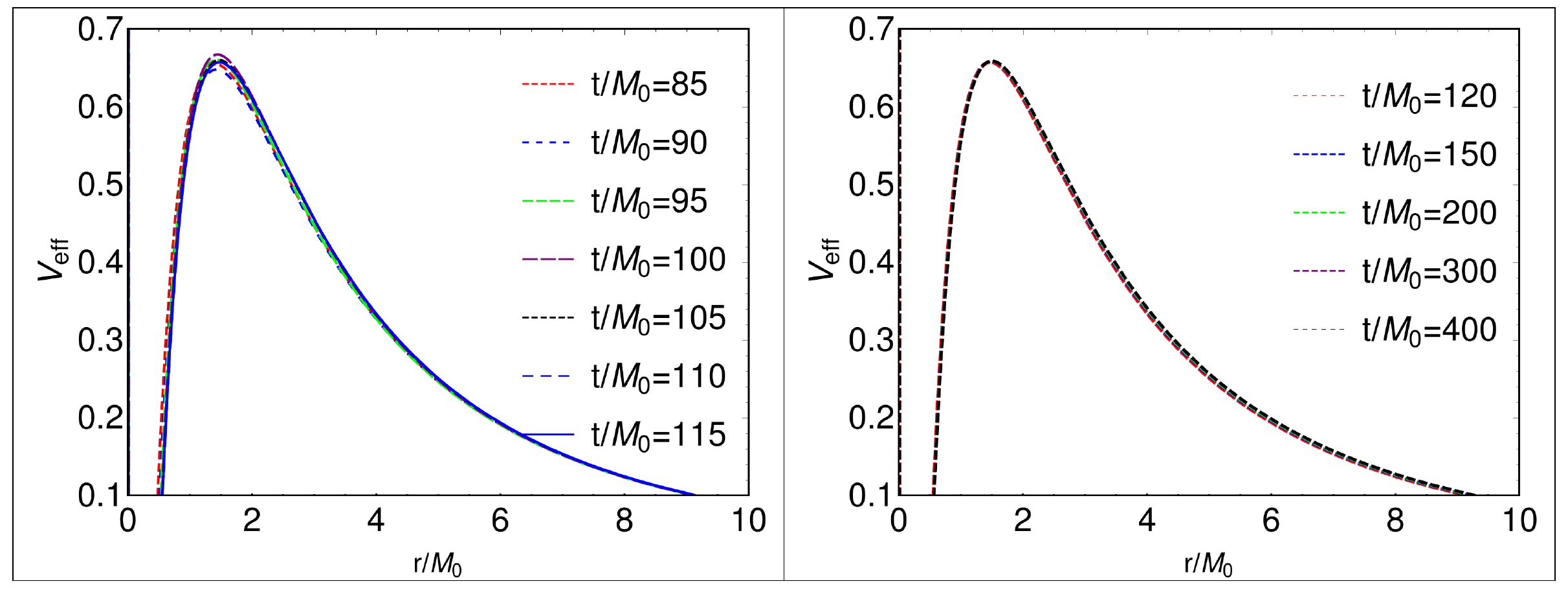}
	\caption{Time evolution of the effective potential for the null geodesics with $\eta=5$ and $L=3.5$.}
	\label{time_veff_null_evo}
\end{figure*}

For simplicity, we only provide the temporal evolution of effective potentials for the case of $\eta=5$, as illustrated in Figs. \ref{time_veff_timelike_evo} and \ref{time_veff_null_evo}, which depict the time-dependent behavior of effective potentials for time-like and null geodesics, respectively. Observing these changes, we find that the onset of dynamical scalarization induces oscillatory behavior in the effective potentials, directly altering their extrema. This implies that tracking such modifications enables one to infer the time evolution of the innermost stable circular orbit (ISCO) radius and the photon sphere radius. After the oscillatory phase subsides, the effective potentials stabilize into nearly stationary configurations as the spacetime asymptotically approaches a almost stationary states at late times.

Having established the time-dependent characteristics of the effective potentials, we proceed to derive the dynamical evolutions of the ISCO and the light ring using these potentials. The areal radius, defined by its geometric significance and its utility in meaningfully comparing radial profiles across solutions \cite{Sanchis-Gual:2021phr}, serves as the appropriate coordinate framework for this analysis. We therefore adopt the areal radius
\begin{equation}
R=r\,e^{2\chi}\sqrt{b}
\end{equation}
to characterize the radii of the ISCO and photon sphere.

\begin{figure*}[htbp]
		\includegraphics[width=\linewidth]{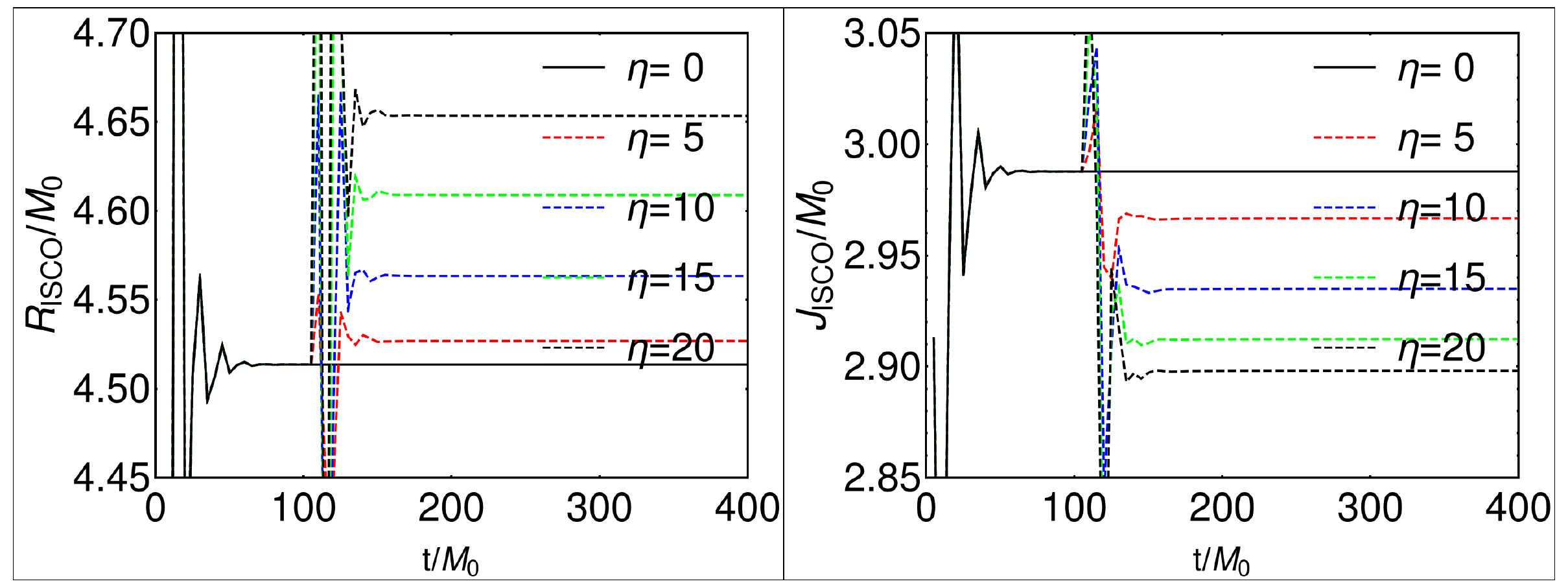}
	\caption{Time evolution of the areal radius and orbital angular momentum of the ISCO in dynamical spacetimes under varying dynamical scalarization with coupling parameter $\eta$.}
	\label{time_ISCO_evo}
\end{figure*}

\begin{figure*}[htbp]
		\includegraphics[width=\linewidth]{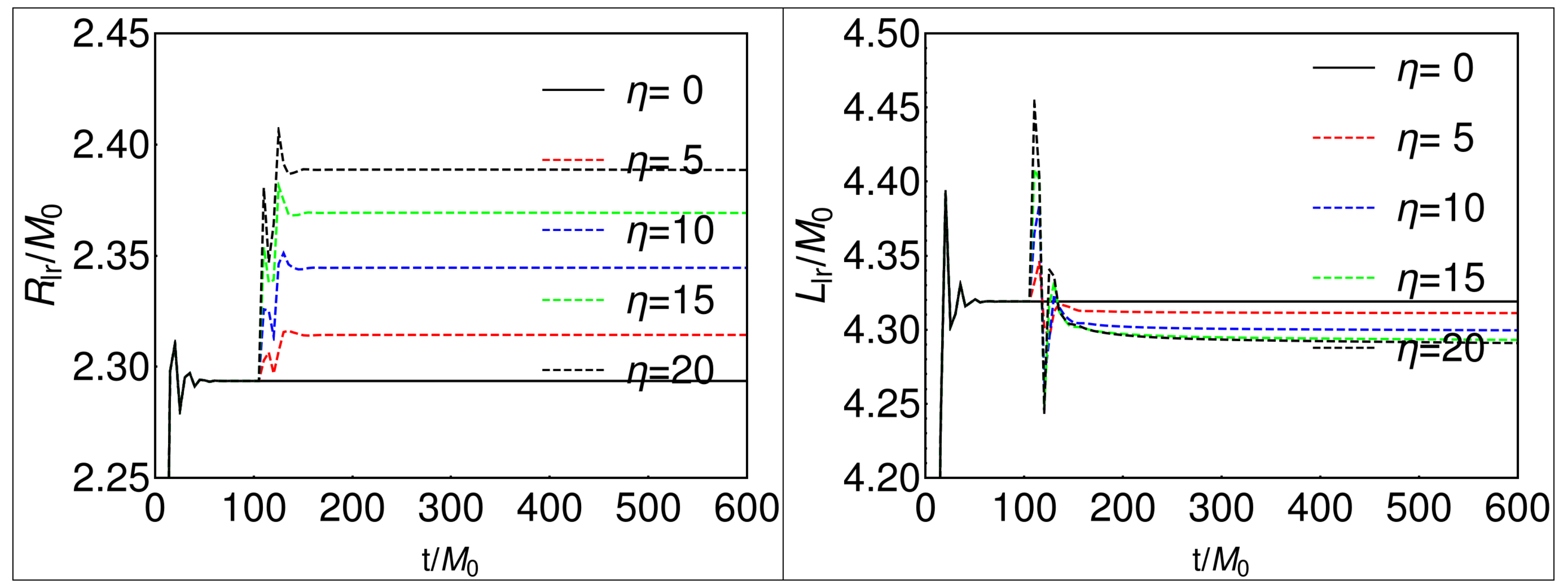}
	\caption{Time evolution of the areal radius and impact parameter of the light ring in the dynamical spacetimes under varying dynamical scalarization with coupling parameter $\eta$.}
	\label{time_lightring_evo}
\end{figure*}

Figures \ref{time_ISCO_evo} and \ref{time_lightring_evo} illustrate the temporal evolution of the areal radii and orbital/angular parameters (orbital angular momentum for the ISCO and impact parameter for the light ring) induced by scalarization. Both systems exhibit oscillatory behavior during their evolution. A comparative analysis of pre- and post-scalarization states reveals that scalarization universally increases the areal radii of both the ISCO and the light ring while decreasing the ISCO's orbital angular momentum and the light ring's impact parameter.

In summary, our numerical simulations of scalarized black holes reveal that the scalarization increases the areal radii of the ISCO and the light ring while reducing their orbital angular momentum (for ISCO) and impact parameter (for light ring). The coupling parameter $\eta$ exhibits a monotonic positive correlation with radii but an inverse relationship with these angular parameters.

\section{Conclusions}\label{Conclusion}
	
We presented the results of studying the time evolution of the ISCO and the light ring induced by dynamical scalarization of a charged black hole. Through nonlinear numerical simulations of the dynamical scalarization process, we derived time-dependent effective potentials at each time step, based on the spacetime metric functions, for both timelike and null geodesics. During the scalarization, energy is transferred from the electromagnetic field to the dilaton scalar field, resulting in an increase in the black hole's irreducible mass and the formation of a surrounding scalar cloud. This energy transfer and scalar cloud formation alter the spacetime geometry, which induces modifications to the effective potentials governing both timelike and null geodesics.

The coordinate transformation occurring during the dynamical evolution of the charged black hole—from the initial isotropic coordinates to the final scalarized configuration—induced changes in the effective potentials \cite{Alcubierre:2002kk}. To clarify how dynamical scalarization impacts the ISCO and light sphere properties, we divided the simulation into two distinct phases. In the first phase, we evolved a purely charged black hole without scalarization, transitioning it from the initial isotropic coordinates to a nearly stationary state in the moving puncture gauge. Once the charged black hole reached this quasi-stationary configuration in the new coordinate frame, we initiated the scalarization process.

We observed significant oscillations in the spacetime’s effective potentials during the initial phase of scalarization, which gradually subsided as the black hole transitioned to a quasi-stationary state. The effective potentials ultimately relaxed to nearly time-independent profiles. To analyze the orbital dynamics under these evolving potentials, we calculated the ISCO and the light ring radii using the time-dependent adiabatic effective potential, tracking their temporal evolution throughout the process. Both the ISCO and light ring exhibited qualitatively similar dynamical behavior, with their properties fluctuating during scalarization before stabilizing. A critical result is that scalarization induced a dual effect: it increased the areal radii of both the ISCO and light ring while simultaneously reducing the orbital angular momentum of the ISCO and the impact parameter of the light ring.

Earlier studies \cite{Mishra:2019trb,Song:2021ziq} investigated the dynamical evolution of the ISCO and the light ring in the analytic Vaidya spacetime, demonstrating that their areal radii increase in the context of an accreting black hole. This radial expansion suggests that black hole scalarization modifies both the inner boundary of accretion disks (defined by the ISCO) and the structure of the black hole shadow (governed by the light ring). Such changes could produce observable signatures, potentially enabling the detection of scalarization effects through astrophysical observations. Our work represented the first study to explicitly examine the dynamical evolution of the ISCO and light ring in the charged black hole regime under scalarization. By focusing on this concrete scenario, we revealed a critical interplay between scalarization-induced spacetime geometry and orbital dynamics. We emphasized that this framework naturally extends to broader dynamical spacetime contexts, such as those driven by superradiant instabilities, accretion processes, or rotational dynamics, offering a versatile tool to probe the dynamics of diverse black hole configurations.

\acknowledgments
	
This work was supported in part by the National Key Research and Development Program of China (Grant No. 2020YFC2201503), the National Natural Science Foundation of China (Grants No. 12105126, No. 12475056, No. 12475055, and No 12247101), the 111 Project under (Grant No. B20063), the Gansu Province's Top Leading Talent Support Plane.


\begin{thebibliography}{100}
		
		\bibitem{Abbott2016a}
		B. P. Abbott \textit{et al.} [LIGO Scientific Collaboration and Virgo Collaboration],
		Phys. Rev. Lett. \textbf{116}, 061102 (2016).
		
		\bibitem{LIGOScientific:2018jsj}
		B.~P.~Abbott \textit{et al.} [LIGO Scientific Collaboration and Virgo Collaboration],
		Astrophys. J. Lett. \textbf{882}, L24 (2019).
		
		\bibitem{LIGOScientific:2020kqk}
		R.~Abbott \textit{et al.} [LIGO Scientific Collaboration and Virgo Collaboration],
		Astrophys. J. Lett. \textbf{913}, L7 (2021).
		
		\bibitem{LIGOScientific:2021psn}
		R.~Abbott \textit{et al.} [LIGO Scientific, VIRGO and KAGRA],
		[arXiv:2111.03634 [astro-ph.HE]].
		
		\bibitem{eth2019}
		K. Akiyama \textit{et al.} [Event Horizon Telescope],
		Astrophys. J. \textbf{875}, L1 (2019); Astrophys. J. \textbf{875}, L2 (2019); Astrophys. J. \textbf{875}, L3 (2019); Astrophys. J. \textbf{875}, L4 (2019); Astrophys. J. \textbf{875}, L5 (2019);Astrophys. J. \textbf{875}, L6 (2019).
		
		\bibitem{eth2022}
		K.~Akiyama \textit{et al.} [Event Horizon Telescope],
		Astrophys. J. Lett. \textbf{930}, L12 (2022); Astrophys. J. Lett. \textbf{930}, L13 (2022); Astrophys. J. Lett. \textbf{930}, L14 (2022); Astrophys. J. Lett. \textbf{930}, L15 (2022); Astrophys. J. Lett. \textbf{930}, L16 (2022); Astrophys. J. Lett. \textbf{930}, L17 (2022).
		
		\bibitem{Chrusciel:2012jk}
		P.~T.~Chrusciel, J.~Lopes Costa, and M.~Heusler,
		Living Rev. Rel. \textbf{15}, 7 (2012).
		
		\bibitem{Chase:1970omy}
		J.~E.~Chase,
		Commun. Math. Phys. \textbf{19}, 276 (1970).
		
		\bibitem{Bekenstein:1971hc}
		J.~D.~Bekenstein,
		Phys. Rev. D \textbf{5}, 1239 (1972).
		
		\bibitem{Bekenstein:1972ky}
		J.~D.~Bekenstein,
		Phys. Rev. D \textbf{5}, 2403 (1972).
		
		\bibitem{Bekenstein:1995un}
		J.~D.~Bekenstein,
		Phys. Rev. D \textbf{51}, R6608 (1995).
		
		\bibitem{Hawking:1972qk}
		S.~W.~Hawking,
		Commun. Math. Phys. \textbf{25}, 167 (1972).
		
		\bibitem{Volkov:1989fi}
		M.~S.~Volkov and D.~V.~Galtsov,
		JETP Lett. \textbf{50}, 346 (1989).
		
		\bibitem{Bizon:1990sr}
		P.~Bizon,
		Phys. Rev. Lett. \textbf{64}, 2844 (1990).
		
		\bibitem{Greene:1992fw}
		B.~R.~Greene, S.~D.~Mathur, and C.~M.~O'Neill,
		Phys. Rev. D \textbf{47}, 2242 (1993).
		
		\bibitem{Luckock:1986tr}
		H.~Luckock and I.~Moss,
		Phys. Lett. B \textbf{176}, 341 (1986).
		
		\bibitem{Droz:1991cx}
		S.~Droz, M.~Heusler, and N.~Straumann,
		Phys. Lett. B \textbf{268}, 371 (1991).
		
		\bibitem{Kanti:1995vq}
		P.~Kanti, N.~E.~Mavromatos, J.~Rizos, K.~Tamvakis, and E.~Winstanley,
		Phys. Rev. D \textbf{54}, 5049 (1996).
		
		\bibitem{Yamazaki:1989hy}
		H.~Yamazaki and I.~Ichinose,
		TIT/HEP-148.
		
		\bibitem{Campbell:1991kz}
		B.~A.~Campbell, N.~Kaloper, and K.~A.~Olive,
		Phys. Lett. B \textbf{285}, 199 (1992).
		
		\bibitem{Garfinkle:1990qj}
		D.~Garfinkle, G.~T.~Horowitz, and A.~Strominger,
		Phys. Rev. D \textbf{43}, 3140 (1991)
		[erratum: Phys. Rev. D \textbf{45}, 3888 (1992)]
		
		\bibitem{Gregory:1992kr}
		R.~Gregory and J.~A.~Harvey,
		Phys. Rev. D \textbf{47}, 2411 (1993).
		
		\bibitem{Guo:2021zed}
		G.-Z.~Guo, P.~Wang, H.-W.~Wu, and H.-T.~Yang,
		Eur. Phys. J. C \textbf{81}, 864 (2021).
		
		\bibitem{Yao:2021zid}
		F.~Yao,
		Eur. Phys. J. C \textbf{81}, 1009 (2021).
		
		\bibitem{Zhang:2021ybj}
		C.-Y.~Zhang, P.~Liu, Y.~Liu, C.~Niu, and B.~Wang,
		Phys. Rev. D \textbf{105}, 024073 (2022).
		
		\bibitem{Zhang:2021nnn}
		C.-Y.~Zhang, Q.~Chen, Y.~Liu, W.-K.~Luo, Y.~Tian, and B.~Wang,
		Phys. Rev. Lett. \textbf{128}, 161105 (2022).
		
        \bibitem{lzqzyb:2023}
        Y. Liu, C.-Y. Zhang, Q. Chen, Z. Cao, Y. Tian, and B. Wang,
        Sci. China-Phys. Mech. Astron. \textbf{66}, 100412 (2023).

		\bibitem{Sotiriou:2013qea}
		T.~P.~Sotiriou and S.-Y.~Zhou,
		Phys. Rev. Lett. \textbf{112}, 251102 (2014).
		
		\bibitem{Sotiriou:2014pfa}
		T.~P.~Sotiriou and S.-Y.~Zhou,
		Phys. Rev. D \textbf{90}, 124063 (2014).
		
		\bibitem{Benkel:2016kcq}
		R.~Benkel, T.~P.~Sotiriou, and H.~Witek,
		Phys. Rev. D \textbf{94}, 121503(R) (2016).
		
		\bibitem{Andreou:2019ikc}
		N.~Andreou, N.~Franchini, G.~Ventagli, and T.~P.~Sotiriou,
		Phys. Rev. D \textbf{99}, 124022 (2019).
		
		\bibitem{Hod:2019pmb}
		S.~Hod,
		Phys. Rev. D \textbf{100}, 064039 (2019).
		
		\bibitem{Peng:2019snv}
		Y.~Peng,
		Nucl. Phys. B \textbf{950}, 114879 (2020).
		
		\bibitem{Ripley:2019aqj}
		J.~L.~Ripley and F.~Pretorius,
		Phys. Rev. D \textbf{101}, 044015 (2020).
		
		\bibitem{Hod:2019vut}
		S.~Hod,
		Eur. Phys. J. C \textbf{79}, 966 (2019).
		
		\bibitem{Liu:2020yqa}
		H.-S.~Liu, H.~Lu, Z.-Y.~Tang, and B.~Wang,
		Phys. Rev. D \textbf{103}, 084043 (2021).
		
		\bibitem{Guo:2020sdu}
		H.~Guo, S.~Kiorpelidi, X.-M.~Kuang, E.~Papantonopoulos, B.~Wang, and J.-P.~Wu,
		Phys. Rev. D \textbf{102}, 084029 (2020).
		
		\bibitem{East:2021bqk}
		W.~E.~East and J.~L.~Ripley,
		Phys. Rev. Lett. \textbf{127}, 101102 (2021).
		
		\bibitem{Zhang:2022kbf}
		Y.-P.~Zhang, Y.-Q.~Wang, S.-W.~Wei, and Y.-X.~Liu,
		Phys. Rev. D \textbf{106}, 024027 (2022).
		
		\bibitem{Tang:2020sjs}
		Z.-Y.~Tang, B.~Wang, T.~Karakasis, and E.~Papantonopoulos,
		Phys. Rev. D \textbf{104}, 064017 (2021).
		
		\bibitem{Silva:2017uqg}
		H.~O.~Silva, J.~Sakstein, L.~Gualtieri, T.~P.~Sotiriou, and E.~Berti,
		Phys. Rev. Lett. \textbf{120}, 131104 (2018).
		
		\bibitem{Doneva:2017bvd}
		D.~D.~Doneva and S.~S.~Yazadjiev,
		Phys. Rev. Lett. \textbf{120}, 131103 (2018).
		
		\bibitem{Dima:2020yac}
		A.~Dima, E.~Barausse, N.~Franchini, and T.~P.~Sotiriou,
		Phys. Rev. Lett. \textbf{125}, 231101 (2020).
		
		\bibitem{Herdeiro:2020wei}
		C.~Herdeiro, E.~Radu, H.~O.~Silva, T.~P.~Sotiriou, and N.~Yunes,
		Phys. Rev. Lett. \textbf{126}, 011103 (2021).
		
		\bibitem{Wang:2020ohb}
		P.~Wang, H.-W.~Wu, and H.-T.~Yang,
		Phys. Rev. D \textbf{103}, 104012 (2021).
		
		\bibitem{Sanchis-Gual:2015lje}
		N.~Sanchis-Gual, J.~C.~Degollado, P.~J.~Montero, J.~A.~Font, and C.~Herdeiro,
		Phys. Rev. Lett. \textbf{116}, 141101 (2016).
		
		\bibitem{Herdeiro:2014goa}
		C.~Herdeiro and E.~Radu,
		Phys. Rev. Lett. \textbf{112}, 221101 (2014).
		
		\bibitem{Press:1972zz}
		W.~H.~Press and S.~A.~Teukolsky,
		Nature \textbf{238}, 211 (1972).
		
		\bibitem{Dolan:2007mj}
		S.~R.~Dolan,
		Phys. Rev. D \textbf{76}, 084001 (2007).
		
		\bibitem{Witek:2012tr}
		H.~Witek, V.~Cardoso, A.~Ishibashi, and U.~Sperhake,
		Phys. Rev. D \textbf{87}, 043513 (2013).
		
		\bibitem{East:2017mrj}
		W.~E.~East,
		Phys. Rev. D \textbf{96}, 024004 (2017).
		
		\bibitem{Bekenstein:1973ur}
		J.~D.~Bekenstein,
		Phys. Rev. D \textbf{7}, 2333 (1973).
		
		\bibitem{Hod:2012wmy}
		S.~Hod,
		Phys. Lett. B \textbf{713}, 505 (2012).
		
		\bibitem{Herdeiro:2018wub}
		C.~Herdeiro, E.~Radu, N.~Sanchis-Gual, and J.~A.~Font,
		Phys. Rev. Lett. \textbf{121}, 101102 (2018).
		
		\bibitem{Fernandes:2019rez}
		P.~G.~S.~Fernandes, C.~Herdeiro, A.~M.~Pombo, E.~Radu, and N.~Sanchis-Gual,
		Class. Quant. Grav. \textbf{36}, 134002 (2019)
		[erratum: Class. Quant. Grav. \textbf{37}, 049501 (2020).]
		
		\bibitem{Berti:2020kgk}
		E.~Berti, L.~G.~Collodel, B.~Kleihaus, and J.~Kunz,
		Phys. Rev. Lett. \textbf{126}, 011104 (2021).
		
		\bibitem{East:2017ovw}
		W.~E.~East and F.~Pretorius,
		Phys. Rev. Lett. \textbf{119}, 041101 (2017).
		
		\bibitem{Julie:2020vov}
		F.~L.~Juli\'e and E.~Berti,
		Phys. Rev. D \textbf{101}, 124045 (2020).
		
		\bibitem{Witek:2020uzz}
		H.~Witek, L.~Gualtieri, and P.~Pani,
		Phys. Rev. D \textbf{101}, 124055 (2020).
		
		\bibitem{East:2020hgw}
		W.~E.~East and J.~L.~Ripley,
		Phys. Rev. D \textbf{103}, 044040 (2021).
		
		\bibitem{Doneva:2021tvn}
		D.~D.~Doneva and S.~S.~Yazadjiev,
		Phys. Rev. D \textbf{105}, L041502 (2022).
		
		\bibitem{Benkel:2016rlz}
		R.~Benkel, T.~P.~Sotiriou, and H.~Witek,
		Class. Quant. Grav. \textbf{34}, 064001 (2017).
		
		\bibitem{Witek:2018dmd}
		H.~Witek, L.~Gualtieri, P.~Pani, and T.~P.~Sotiriou,
		Phys. Rev. D \textbf{99}, 064035 (2019).
		
		\bibitem{Doneva:2020nbb}
		D.~D.~Doneva, L.~G.~Collodel, C.~J.~Kr\"uger, and S.~S.~Yazadjiev,
		Phys. Rev. D \textbf{102}, 104027 (2020).
		
		\bibitem{Doneva:2021dcc}
		D.~D.~Doneva and S.~S.~Yazadjiev,
		Phys. Rev. D \textbf{103}, 083007 (2021).
		
		\bibitem{Doneva:2021dqn}
		D.~D.~Doneva and S.~S.~Yazadjiev,
		Phys. Rev. D \textbf{103}, 064024 (2021).
		
		\bibitem{Abramowicz:2011xu}
		M.~A.~Abramowicz and P.~C.~Fragile,
		Living Rev. Rel. \textbf{16}, 1 (2013).
		
		\bibitem{Cardoso:2008bp}
		V.~Cardoso, A.~S.~Miranda, E.~Berti, H.~Witek, and V.~T.~Zanchin,
		Phys. Rev. D \textbf{79}, 064016 (2009).
		
		\bibitem{Luminet:1979nyg}
		J.~P.~Luminet,
		Astron. Astrophys. \textbf{75}, 228 (1979).
		
		\bibitem{Falcke:1999pj}
		H.~Falcke, F.~Melia, and E.~Agol,
		Astrophys. J. Lett. \textbf{528}, L13 (2000).
		
		\bibitem{Gralla:2019xty}
		S.~E.~Gralla, D.~E.~Holz, and R.~M.~Wald,
		Phys. Rev. D \textbf{100}, 024018 (2019).
		
		\bibitem{Mishra:2019trb}
		A.~K.~Mishra, S.~Chakraborty, and S.~Sarkar,
		Phys. Rev. D \textbf{99}, 104080 (2019).
		
		\bibitem{Song:2021ziq}
		Y.~Song,
		Eur. Phys. J. C \textbf{81}, 875 (2021).
		
		\bibitem{Zhang:2023qag}
		Y.-P.~Zhang, S.-X.~Sun, Y.-Q.~Wang, S.-W.~Wei, P.~Laguna, and Y.-X.~Liu,
		Phys. Rev. Res. \textbf{6}, 033187 (2024).
		
		\bibitem{Cunha:2022gde}
		P.~V.~P.~Cunha, C.~Herdeiro, E.~Radu, and N.~Sanchis-Gual,
		Phys. Rev. Lett. \textbf{130}, 061401 (2023).

        \bibitem{Zhang:2024wci}
        Y.-P.~Zhang, S.-J.~Yang, S.-W.~Wei, W.-D.~Guo, and Y.-X.~Liu,
        Phys. Rev. D \textbf{111}, 104005 (2025).

        \bibitem{Montero:2012yr}
		P.~J.~Montero and I.~Cordero-Carrion,
		Phys. Rev. D \textbf{85}, 124037 (2012).
	
		\bibitem{Alcubierre:2009ij}
		M.~Alcubierre, J.~C.~Degollado, and M.~Salgado,
		Phys. Rev. D \textbf{80}, 104022 (2009).
		
		\bibitem{Alcubierre:2011pkc}
		M.~Alcubierre and M.~D.~Mendez,
		Gen. Rel. Grav. \textbf{43}, 2769 (2011).
		
		\bibitem{Sanchis-Gual:2021phr}
		N.~Sanchis-Gual, C.~Herdeiro, and E.~Radu,
		Class. Quant. Grav. \textbf{39}, 064001 (2022).

		\bibitem{Alcubierre:2002kk}
		M.~Alcubierre, B.~Bruegmann, P.~Diener, M.~Koppitz, D.~Pollney, E.~Seidel, and R.~Takahashi,
		Phys. Rev. D \textbf{67}, 084023 (2003).
		
	\end{thebibliography}
\end{document}